# Suppression of the spin waves non-reciprocity due to interfacial Dzyaloshinskii–Moriya interaction by lateral confinement in magnetic nanostructures

S. Tacchi[1], R. Silvani[2], M. Kuepferling[3], A. Fernández Scarioni[4], S. Sievers[4], H.W. Schumacher[4], E. Darwin[5,6], M.-A. Syskaki[7,8], G. Jakob[7], M. Kläui[7] and G. Carlotti[2,1]

[1] *Istituto Officina dei Materiali del CNR (IOM-CNR), c/o Department of Physics and Geology, University of Perugia, Italy*
[2] *Department of Physics and Geology, University of Perugia, Italy*
[3] *Istituto Nazionale di Ricerca Metrologica, Torino, Italy*
[4] *Physikalisch-Technische Bundesanstalt, Braunschweig, Germany*
[5] *School of Physics and Astronomy, University of Leeds, United Kingdom*
[6] *Department of Electrical and Information Engineering, Politecnico di Bari, Italy*
[7] *Institute of Physics, Johannes Gutenberg-University Mainz, Mainz, Germany*
[8] *Singulus Technologies AG, Kahl, Germany*

**ABSTRACT**

Despite the huge recent interest towards chiral magnetism related to the interfacial Dzyaloshinskii–Moriya interaction (i-DMI) in layered systems, there is a lack of experimental data on the effect of i-DMI on the spin waves eigenmodes of laterally confined nanostructures. Here we exploit Brillouin Light Scattering (BLS) to analyze the spin wave eigenmodes of non-interacting circular and elliptical dots, as well as of long stripes, patterned starting from a Pt(3.4 nm)/CoFeB(0.8 nm) bilayer, with lateral dimensions ranging from 100 nm to 400 nm. Our experimental results, corroborated by micromagnetic simulations based on the GPU-accelerated MuMax3 software package, provide evidence for a strong suppression of the frequency asymmetry $\Delta f$ between counter-propagating spin waves (corresponding to either Stokes or anti-Stokes peaks in BLS spectra), when the lateral confinement is reduced from 400 nm to 100 nm, i.e. when it becomes lower than the light wavelength. Such an evolution reflects the modification of the spin-wave character from propagating to stationary and indicates that the BLS-based method of quantifying the i-DMI strength from the frequency difference of counter-propagating spin waves is not applicable in the case of magnetic elements with lateral dimension below about 400 nm.

## I. INTRODUCTION

A renewed interest for chiral magnetism and its possible applications has been triggered by the discovery of the interfacial Dzyaloshinskii–Moriya interaction (i-DMI) in magnetic thin films [1,2,3]. This is the antisymmetric exchange interaction arising thanks to spin-orbit coupling at the interface between a ferromagnetic film and a heavy-metal substrate [4,5,6,7]. One of the main consequences of the presence of i-DMI is the marked non-reciprocity in the spin waves (SWs) propagation [8,9,10,11] and this can be exploited to quantify the i-DMI strength, $D$, in continuous films and multilayers. To this respect, some coauthors of this paper have presented very recently a comprehensive review [12] about the methods for the determination of the DMI strength, showing that the most accurate and reliable method is the Brillouin light scattering (BLS) technique. This is based on the measurement of the frequency difference of oppositely propagating spin waves, that appear on the Stokes and anti-Stokes sides of BLS spectra [13, 14, 15, 16, 17, 18, 19, 20]. From an applications point of view, such a non-reciprocity can be useful to design one-way devices, such as SW diodes, isolators or circulators, necessary to proceed towards the developing of nano-magnonic circuits, operating in the frequency range 1-100 GHz, that are crucial for information and communication technology [21,22,23,24,25]. Moreover, in the emerging field of magnon spintronics, i-DMI has been theoretically proposed as an additional means to tailor the magnonic band structure of chiral magnonic crystals (MCs), where an artificial periodic modulation of the magnetic properties is introduced by nanofabrication. In particular, it has been predicted that the presence of the i-DMI in MCs causes a non-reciprocity of the dispersion curves and a modification of both the frequency and the spatial profiles of the allowed magnonic modes [26, 27, 28, 29].

In contrast with the huge number of both theoretical and experimental results relative to extended films and multilayers [12], only a few theoretical investigations have been devoted to artificial systems characterized by a laterally confined geometry, such as magnetic dots or wires, while experimental investigations are still lacking in literature. In a pioneering micromagnetic study, Garcia-Sanchez et al. [30] showed theoretically that Néel domain walls induced by the i-DMI can modify spin-wave propagation by producing a non-reciprocal channeling along the center of the wall. In addition, it is expected that i-DMI leads to a large frequency splitting of eigenmodes with a strong azimuthal character in perpendicularly magnetized dots. Subsequently, M. Mruczkiewicz et al. [31] exploited both the frequency-domain method and micromagnetic simulations to investigate the impact of i-DMI on the ferromagnetic resonance (FMR) spectrum of isolated stripes magnetized in-plane, where i-DMI causes a red-shift of some peaks in the simulated FMR spectrum, as well as the appearance of new peaks. More recently, two theoretical investigations of nanodots magnetized in-plane were presented by Zingsem et al. [32] and by some of the coauthors of the present paper [33]. The former paper proposed a model for quantized traveling waves in a confined geometry for the case of non-reciprocal propagation, providing evidence that confinement in a non-reciprocal energy landscape produces a fixed nodal structure with wave amplitudes modulated in time and a phase velocity determined by the difference in wavelengths between opposite propagation directions. In the latter paper, it was shown that the calculated eigenmodes spectrum of elliptical nanodots is appreciably modified by the i-DMI induced non-reciprocity: the frequencies of the eigenmodes are red-shifted and their spatial profiles appreciably altered due to the lack of stationary character in the direction orthogonal to the magnetization direction.

In the present work we aim to fill the lack of experimental data relative to laterally confined nanosystems, exploiting the BLS technique to provide experimental evidence of the effect of i-DMI on the SW eigenmodes of magnetic wires and dots magnetized in-plane. We analyze the eigenmodes of non-interacting circular and elliptical dots, as well as of long stripes, patterned starting from a Pt(3.4 nm)/CoFeB(0.8 nm) bilayer, with lateral dimensions ranging from 100 to 400 nm, i.e. comparable with the wavelength of the thermal spin waves detected in BLS experiments. Micromagnetic simulations are performed to shed light on the interpretation of the experimental results.

## II. EXPERIMENTAL
### A) Sample preparation

The thin film material stack, substrate/Ta(5.7)/Pt(3.4)/ CoFeB (0.8)/MgO(1.4)/Ta(5) (thickness unit: nm), was deposited on Si/SiO$_2$ substrate, with the Singulus Rotaris magnetron sputtering tool at a base pressure of $5\times10^{-8}$ mbar. DC-magnetron sputtering was employed at room temperature for the growth of the metallic layers Ta, Pt and Co$_{60}$Fe$_{20}$B$_{20}$(CoFeB), whereas RF-sputtering was used for the growth of the oxide layer from a composite target (MgO). The deposition rates for Ta, Pt, CoFeB and MgO were 0.027, 0.094, 0.135 and 0.008 nm/s, respectively, under a pure Ar flow used as sputtering gas. The top Ta layer serves as capping for the material stack to avoid oxidation over patterning and time. The saturation magnetization was measured via SQUID magnetometry. The area of the sample was found by photographing it next to a scale and calculating the mm/pixel ratio, and the ferromagnet thickness used was $t$=0.8 nm. An in-plane SQUID measurement was taken, a filling factor was applied, and the value of the saturation magnetization was determined to be $M_s=1.4 \pm 0.1$ MA/m. The different structures on the measured sample were first defined with an electron beam writer on a 140 nm thick PMMA (polymethyl methacrylate) resist layer. The PMMA resist was developed, and a 50 nm thick aluminum layer was evaporated; immediately afterwards a resist lift-off process left only the desired aluminum patterns on the multilayer film. The structures were patterned by etching with an ion source using argon, with the aluminum patterns serving as a hard mask. Finally, the aluminum layer was chemically removed with an aqueous NaOH-based alkaline developer, leaving only the structures consisting of the multilayer film. In Fig. 1, top panels, scanning electron microscopy images of three samples are shown.

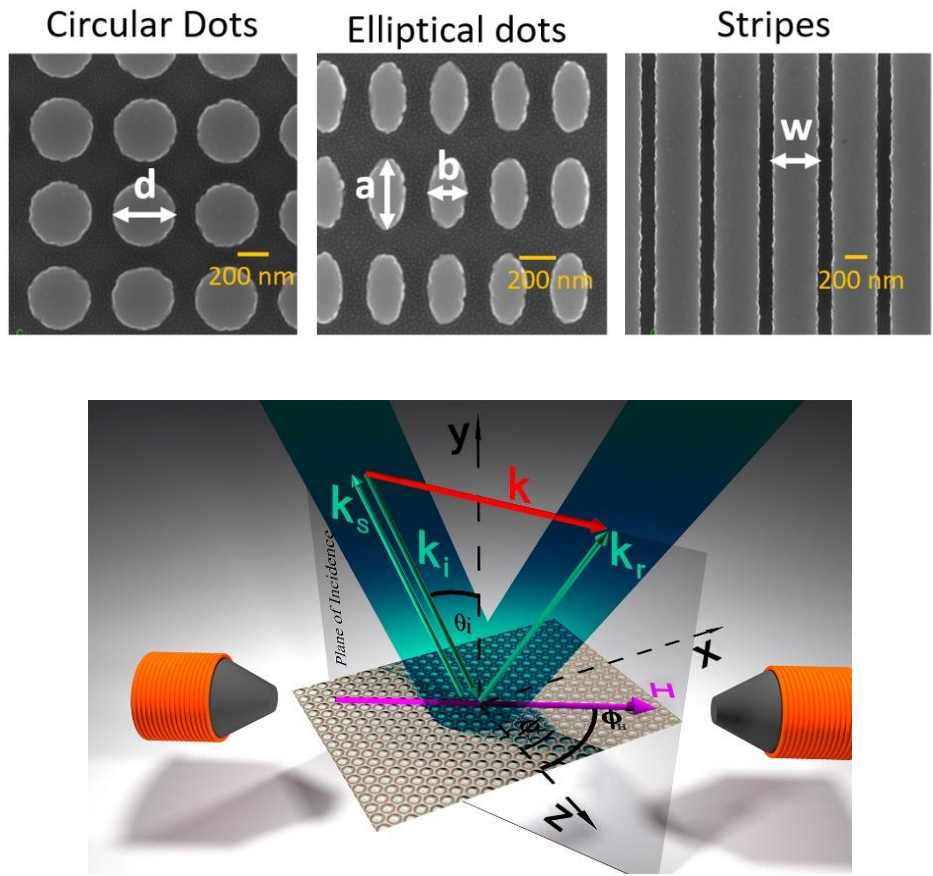

**Fig. 1** Upper panels: scanning electron microscope images of some of the magnetic nanoelements studied in this work: circular dots, elliptical dots and long wires. The yellow bar is 200 nm. Bottom panel: sketch of the wavevector-resolved BLS experiment: $k_i$, $k_s$ and $k_r$ represent the wavevectors of the incoming, scattered and reflected light, respectively, while $k$ represent the wavevector of spin waves. H is the in-plane applied field, that in the experiment was applied along the z-axis ($\phi_H=0°$ for $H>0$ and $\phi_H=180°$ for $H<0$), while the in-plane wave vector $\boldsymbol{k}$ was swept along the perpendicular direction ($\phi_k=90°$), in the so-called Damon-Eshbach (DE) configuration.

### B) BLS experiments

BLS measurements were performed by focusing about 200 mW of a monochromatic laser beam ($\lambda$=532 nm) onto the sample surface, as sketched in Fig. 1, bottom panel, where the reference frame, the involved wavevectors and the relevant angles are also indicated. The illuminated area on the sample, thanks to the focusing lens (a camera objective with a 50 mm focal length), was a circle of approximately 40 µm in diameter, so thousands of nanoelements were probed simultaneously. The back-scattered light was analyzed by a Sandercock-type (3+3)-pass tandem Fabry-Perot interferometer. An in-plane magnetic field $H$ was applied parallel to the film plane, along the z-axis ($\phi_H$=0° for $H$>0 and $\phi_H$=180° for $H$<0), while the in-plane wave vector $k$ was swept along the perpendicular direction ($\phi_k$=90°), in the so-called Damon-Eshbach (DE) configuration, as shown in the bottom panel of Fig. 1. Due to the conservation of momentum in the light scattering process, the magnitude of $k$ is related to the incidence angle of light $\theta_i$, by the relation $k=4\pi \sin\theta_i/\lambda$ and spans from zero to about 20 rad/µm.

### III. THEORETICAL FRAMEWORK AND MICROMAGNETIC SIMULATIONS

The dynamics of the considered magnetic systems are governed by the well-known Landau-Lifshitz-Gilbert equation (LLG)[12]:

$$\frac{\partial \boldsymbol{m}}{\partial t} = -\gamma \left(\boldsymbol{m} \times \boldsymbol{H}_{eff}\right) + \alpha \left(\boldsymbol{m} \times \frac{\partial \boldsymbol{m}}{\partial t}\right) \qquad (1)$$

where $\boldsymbol{m}(\boldsymbol{r},t) = \frac{\boldsymbol{M}}{M_s}$ is the unit vector along the local magnetization, $\gamma$ is the gyromagnetic ratio and $\alpha$ is the Gilbert damping constant. $\boldsymbol{H}_{eff}$ is the effective field acting on the processing spins, which can be derived from the energy functional $E_{tot}$ and consists of several contributions, reflecting the different energy terms, according to:

$$\boldsymbol{H}_{eff} = -\frac{1}{\mu_0}\frac{\partial E_{tot}}{\partial \boldsymbol{m}} = \boldsymbol{H} + \boldsymbol{H}_{exch} + \boldsymbol{H}_{ani} + \boldsymbol{H}_{ms} + \boldsymbol{H}_{DMI} \qquad (2)$$

where $\boldsymbol{H}$ is the external applied field, $\boldsymbol{H}_{exch} = \frac{2A}{\mu_0 M_s}\nabla^2 \boldsymbol{m} = J\nabla^2 \boldsymbol{m}$ is the exchange field and $A$ the exchange stiffness constant, $\boldsymbol{H}_{ani} = \frac{2K_u}{\mu_0 M_s}m_y\widehat{e_y}$ is the anisotropy field with $K_u$ the uniaxial perpendicular anisotropy constant, and $\boldsymbol{H}_{DMI}=\frac{2D}{\mu_0 M_s}\left(\frac{\partial m_y}{\partial x}\widehat{e_x}-\frac{\partial m_x}{\partial x}\widehat{e_y}\right)$ with $D$ the DMI constant. The magnetostatic dipolar field $\boldsymbol{H}_{ms}$ would be uniform in the case of an ellipsoidal body only, so for planar nanostructures it is usually calculated numerically, accounting for its non-uniformity and non-locality.

In the case of ultrathin films it is possible to achieve a simple analytical expression of the dispersion curve, as follows [12]:

$$f(k) = f_0(k) \pm f_{DMI}(k) =$$
$$= \gamma\mu_0\sqrt{(H+Jk^2+P(kt)M_s)\left(H+Jk^2+M_s-H_{ani}-P(kt)M_s\right)} \pm \frac{\gamma D}{\pi M_s}k \quad (3)$$

where the dipolar term $P(kt) = 1 - \frac{1-e^{-|kt|}}{|kt|}$ that reduces to $|kt|/2$ in the case of ultrathin films, being $t$ the ferromagnetic film thickness. This means that the frequency difference between the Stokes and anti-Stokes peaks measured in BLS spectra corresponds to:

$$\Delta f = 2f_{DMI} = \frac{2\gamma D}{\pi M_s}k \quad (4)$$

Note that the sign of $\Delta f$ and $D$, derived from the experiment, depends on the convention chosen in both the definition of the i-DMI Hamiltonian and the geometry of interaction [12]. In our choice, a positive $D$ corresponds to a favored right-handed (clockwise) chirality (↑→↓ or ↓←↑), while a negative $D$ corresponds to a favored left-handed (counterclockwise) chirality (↑←↓ or ↓→↑).

The most general methods to solve the LLG equation above are based on proper numerical techniques where it is spatially discretized using finite differences or finite elements methods. As a result, a discretized version of the effective field is obtained and the corresponding system of differential equations are solved within suitable time-stepping schemes. The Micromagnetic simulations were performed by the GPU-accelerated software MuMax3 [34].

In order to mimic the round shape of the dot borders, the lateral size of the discrete cells was taken as small as $1 \times 1$ nm², while the cell thickness was the same as the reference film, i.e. $t=0.8$ nm. In the case of the stripes, in order to reproduce their infinite length, the periodic boundary conditions were applied in the direction parallel to the stripes. Then, the magnetization dynamics were excited using a temporal sinc-shaped field pulse, with an amplitude of $\mu_0 H_y = 10$ mT, directed perpendicular to the sample plane (along the $y$ axis), able to activate modes up 25 GHz. To better excite modes with different spatial profiles, the pulse was localized in a region corresponding to approximately one third of the element width from the left edge and the temporal evolution of the magnetization dynamics was recorded for a total time of 20 ns (10 ns in the simpler case of the extended film). As it will be discussed further on, in the case of the stripes, the exciting field pulse was also localized in a 5 nm narrow region close to the edge, to better excite the low-frequency mode that was observed in the experiments. The spatiotemporal fast Fourier-transform (FFT) of the recorded evolution of the out-of-plane magnetization of the whole set of discrete cells was then performed to obtain the dispersion curves (frequency vs. $k$) of the magnonic modes. The Gilbert damping was fixed at $\alpha = 0.001$, to obtain a narrow linewidth of the peaks in the eigenmodes spectra, while the other magnetic parameters were obtained from a careful analysis of the BLS measured frequencies in the reference extended film, as detailed in the next section.

## IV. RESULTS AND DISCUSSION
### A) Extended film

As a first step of our investigation, we have carried on BLS measurements on the extended film, to quantify the value of the DMI constant $D$. In Fig. 2 we present the measured frequencies (open dots) for both positive and negative wavenumbers (corresponding to the Stokes and anti-Stokes peaks measured in the BLS spectra, respectively). It is clearly seen that the dispersion curve shows a marked non-reciprocal behavior. The uniaxial out-of-plane anisotropy constant $K_u$=1.09 MJ/m$^3$ and the gyromagnetic ratio $\gamma$=1.85×10$^{11}$ rad/Ts were obtained from a best fit procedure of the average frequency $f_0(k)$ measured as a function of the intensity of the external field. This analysis was carried out fixing $M_S$ to the value measured by SQUID magnetometry and the exchange constant $A$ to the literature value of $A$=10 pJ/m (note that the latter parameter does not appreciably influence the SW frequency for films of such small thickness). Then, the value of $D$ was obtained from the analysis of the SW dispersion, measured for $\mu_0H = \pm 200$ mT, using Eq. 3. As one can see in Fig. 2 the measured non-reciprocity could be nicely reproduced by the analytical formula (yellow curve) assuming an interfacial Dzyaloshinskii-Moriya strength $D$=1.48 mJ/m$^2$ with an uncertainty of about 10%. Moreover, we found a good agreement between the experimental results and the intensity of the perpendicular component of the dynamical magnetization obtained by the micromagnetic simulations (grey-scale curve in Fig. 2) performed using the magnetic parameters obtained from the BLS analysis. With such an information about the magnetic properties at hands, we proceeded with the analysis of the patterned samples.

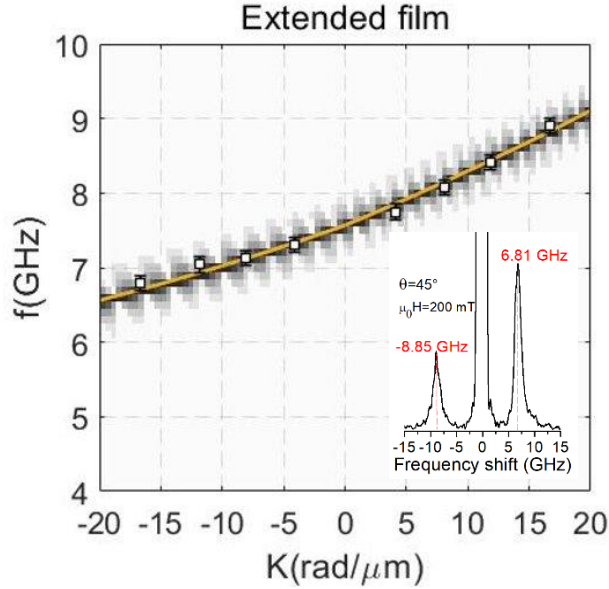

**Fig. 2** Open dots: measured frequencies of the Stokes (positive *k*) and anti-Stokes (negative *k* ) peaks in the BLS spectra, for an applied field $\mu_0H$=200 mT. The yellow curve represents the dispersion curve calculated according to Eq. 3, while the gray-scale curve is the result of micromagnetic simulation. The best-fit value of the DMI constant, responsible for the non-reciprocal character of this dispersion curve, is *D*=1.48 mJ/m$^2$. The inset shows a typical BLS spectrum.

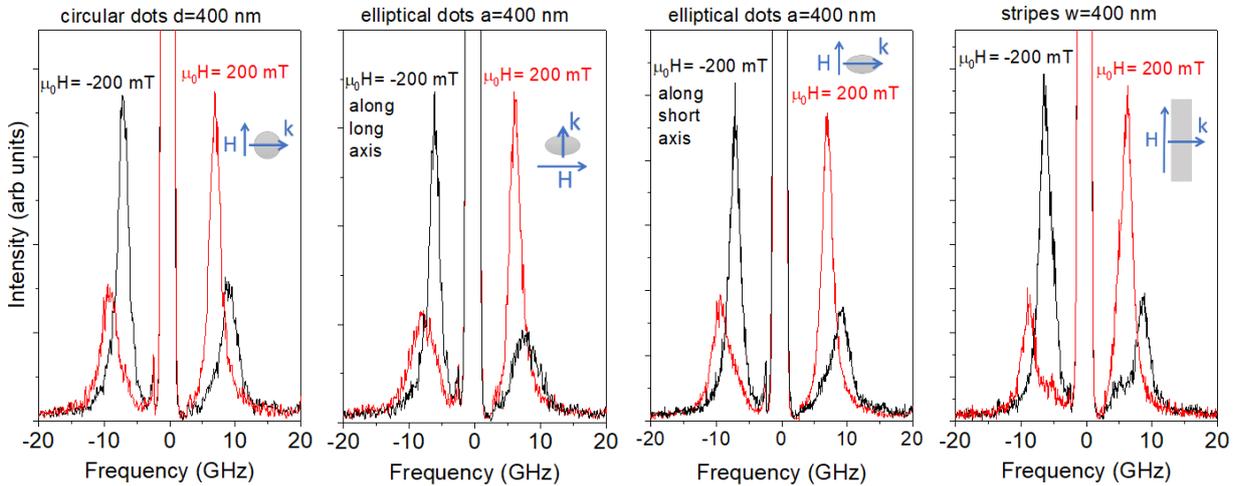

Fig. 3 Typical BLS spectra recorded from the patterned samples with dots or wires of lateral dimension 400 nm, for negative and positive applied field *H*=±200 mT, at an incidence angle $\theta_i$=45°. It is evident the DMI-induced non-reciprocity of the BLS spectra when the direction of the applied field H is reversed. The small insets indicate the adopted interaction geometry.

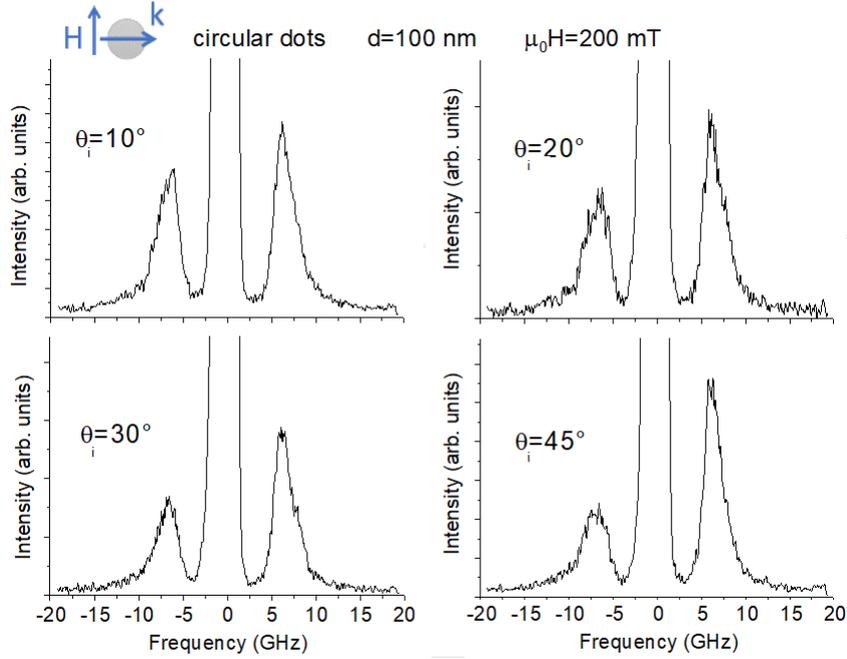

Fig. 4 BLS spectra recorded from the patterned samples with circular dots of diameter d=100 nm, for different angles of incidence $\theta_i$.

**B) Patterned samples**

In the second step of our investigation, we have analyzed the patterned samples containing circular and elliptical dots, as well as long stripes, having lateral dimensions ranging from 100 to 400 nm. On the basis of preliminary micromagnetic simulations and of previous investigations of dense matrices of magnetic dots [35], the edge-to-edge distance between adjacent elements was designed to be about 150 nm (and found to be between 130 nm and 180 nm in real samples), that is sufficient to consider them as isolated elements, taking into account the sub-nanometric thickness of the magnetic elements. Fig. 3 show typical spectra recorded for positive and negative field $\mu_0 H = \pm 200$ mT and an angle of incidence $\theta_i$=45° for the dots and wires with lateral dimension 400 nm. It is seen that in all the cases we could put in evidence a remarkable non-reciprocity of the BLS spectra, as expected in the presence of DMI. The evolution of such non-reciprocity was then analyzed as a function of the angle of incidence $\theta_i$, i.e of the spin wave wavevector, as seen in Fig. 4 for the circular dots of diameter d= 100 nm. Please note that the slightly different lineshape observed for the Stokes and the antiStokes peaks can be attributed to the presence of DMI, as found in previous investigations of extended Co or CoFeB, because the DMI-induced term is antisymmetric in the wave vector [36,37]. Now we discuss in details the results obtained for the different class of patterned samples, starting from the circular dots.

### B1) Circular dots

BLS spectra were recorded at four different angles of incidence $\theta_i=10°$, $20°$, $30°$ and $45°$, corresponding to four distinct values of the SW wavenumber $k=4\pi \sin\theta_i/\lambda$. As shown in the left panel of Fig. 3 and in the different panels of Fig. 4, the measured spectra contain only one peak with a pronounced linewidth, whose position changes with the angle of incidence, i.e. with $k$, as represented by the full dots in the three panels of Fig. 5(b,c,d). Again, positive and negative wavenumbers correspond to the Stokes and anti-Stokes peaks measured in the BLS spectra, respectively. From a comparison of the measured frequencies with the dispersion curves of the extended film (yellow curves) it is clear that only for the dots with the largest diameter $d=400$ nm, the measured points are very close to the dispersion curve of the extended film. Moreover, the observed frequency asymmetry $\Delta f$ between the frequencies corresponding to positive and negative wavevectors coincides, within the experimental error, with those of the extended film, as seen in Fig. 5(a). On the contrary, for the smaller dots, with $d=200$ and $100$ nm, the observed frequency asymmetry $\Delta f$ is appreciably suppressed, if compared to the case of the extended film. Therefore, if one derives the value of the "effective" i-DMI constant using Eq. 4, as in the case of extended films, one would obtain i-DMI values appreciably reduced as the dots diameter reduces below 400 nm. Quantitatively, the obtained values of the "effective" i-DMI constant $D_{eff}$, that are indicated in Fig. 5(a), have an uncertainty of about 10%, account for a reduction of about 40% (for $d=200$ nm) and 67% (for $d=100$ nm), with respect to the continuous film. To shed light on this behavior, let us consider the results of the micromagnetic simulations, that are visible as grey-scale curves in the three panels of Fig. 5(b,c,d). In the simulations, one can observe a number of discrete frequencies, corresponding to the SW eigenmodes, which become denser upon increasing the dot diameter. The fact that in the experiment we detected only one broad peak and we could not resolve the individual discrete modes obtained in the simulations, can be explained observing that the separation between the simulated frequencies of adjacent modes is below one GHz, even for the sample with the smallest dots (Fig. 5(b)). This frequency difference is too small to be resolved in the experiment for at least two concomitant reasons. First of all, the damping coefficient of the experimentally used CoFeB nanoelements is expected to be more than an order of magnitude larger than the value α=0.001 assumed in the simulations, causing a broadening of the BLS peaks.

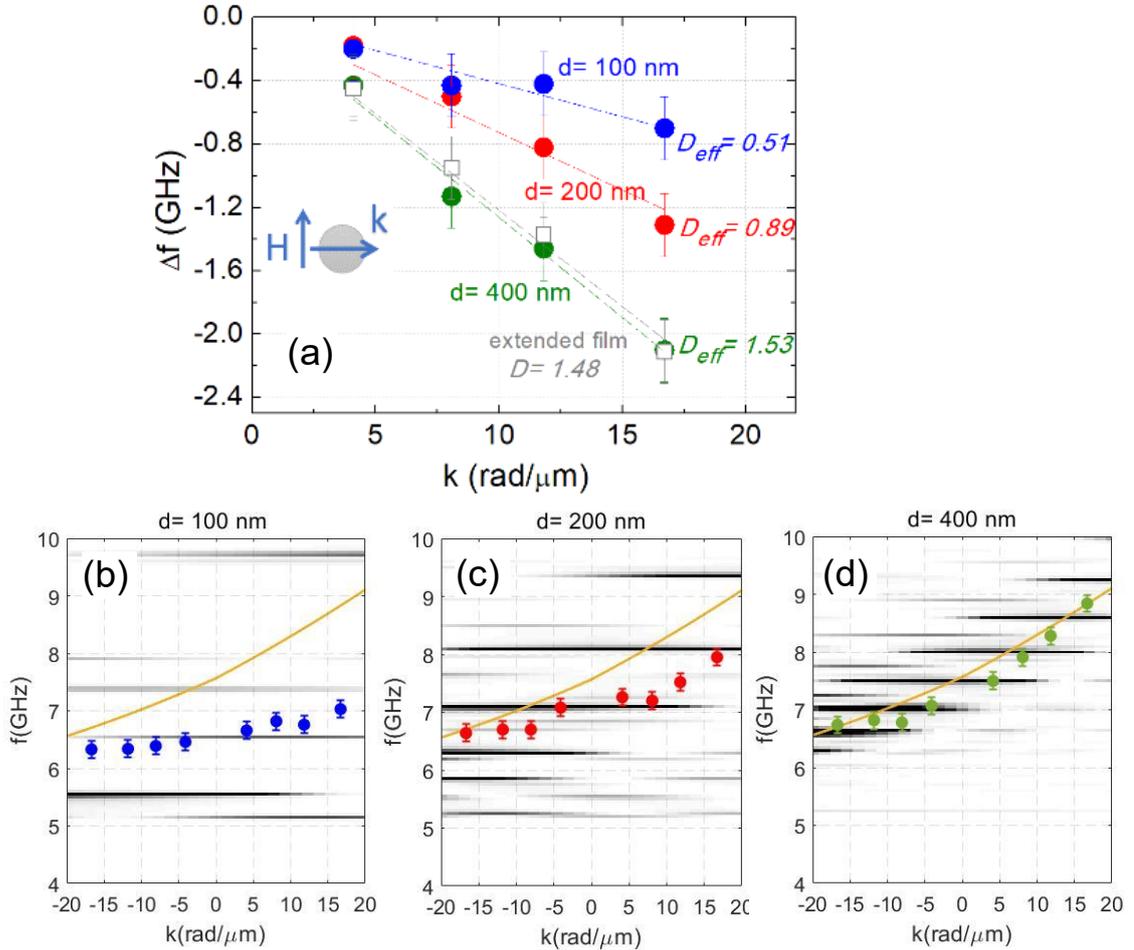

**Fig.** 5 (a) Dependence of the measured frequency asymmetry $\Delta f$ between the Stokes and the anti-Stokes peaks in BLS spectra for the three samples of circular dots of diameter d (closed circles) and for the extended film (open squares). The slope of the regression lines (dashed lines) is proportional, according to Eq. 4, to the effective *i-DMI* constant $D_{eff}$ whose values are expressed in units of mJ/m$^2$; it is seen that the value of such $D_{eff}$ is appreciably reduced, with respect to the value of $D$ in the extended film, for $d$=200 nm and 100 nm. (b), (c), (d) Measured (full circles) and simulated (grey-scale curves) eigenmode frequencies for the circular dots of diameter $d$=100, 200 and 400 nm, respectively. It can be seen that the simulated eigenmodes are more dense in frequency as the dot diameter is increased from $d$=100 nm to 400 nm, while the extension in the k space of each mode tends to be reduced and concentrated close to the dispersion curve of the extended film (yellow line), reflecting the conservation of momentum that becomes effective as the lateral dimension of the dots becomes comparable with the SW wavelength.

Secondly, as anticipated in the experimental section, the measured signal comes from the illuminated region of the sample (a circle of about 40 μm in diameter for conventional BLS measurements) that involves thousands of nanoelements which, in practice, present differences and imperfections leading to a further inhomogeneous broadening of the BLS peaks. For the above reasons, the linewidth of the experimental peak is of 2-3 GHz, as seen in Figs. 3 and 4, so that adjacent discrete modes may contribute to the experimental peak, although they could not be resolved. Finally, it is interesting to note in Fig. 5(b,c,d) that, thanks to the presence of i-DMI, the simulated modes do not have a simple stationary character, but they exhibit a propagative character along the direction of the exchanged wavevector *k*. As a consequence, one can see that the intensity of the modes evolves, shifting to higher frequency, when one moves from negative to positive values of *k*. In particular, for the largest dot diameter *d*=400 nm, the eigenmode intensity follows rather well the dispersion curve of the extended film, as found in the experiment (green points). This is due to the fact that for *d*=400 nm the discrete spectrum is sufficiently dense and the range of *k* where each mode has a large intensity is relatively restricted. On the contrary, for *d*=100 nm the discrete modes are far apart in frequency and their intensity is almost constant with varying *k* (because the conservation of momentum does not apply when the lateral dimension of the dot is much smaller than the SW wavelength, a manifestation of the exclusion principle). As a consequence, the main intensity remains in the two low-lying discrete modes and in fact the experimental points do not follow the relatively large slope of the dispersion curve of the extended film. An intermediate situation is found for *d*=200 nm.

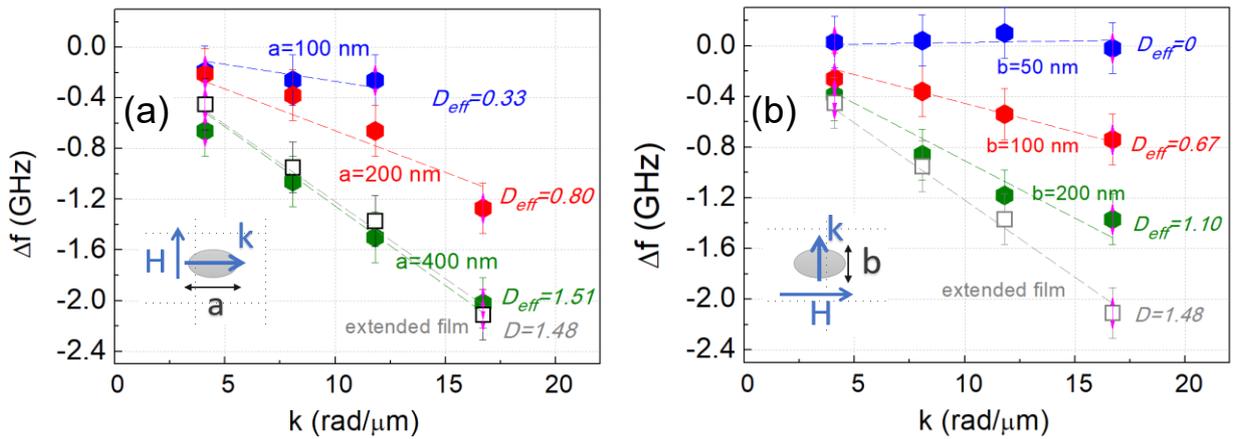

**Fig. 6** Dependence of the measured frequency asymmetry *Δf* between the Stokes and the anti-Stokes peaks in BLS spectra, for the three samples of elliptical dots (full hexagons) and for the extended film (open squares), for SW wavevector *k* directed parallel to the long (a) and short (b) axis of the elliptical dots, respectively, as sketched in the insets. The slope of the regression lines is proportional to the effective i-DMI constant $D_{eff}$, according to Eq. 4, whose values are expressed in $mJ/m^2$. It is seen that the value of the effective *i-DMI* constant extracted using Eq. 4 is appreciably reduced, compared to the value of the extended film, when the lateral confinement for SW propagation is below 400 nm.

### B2) Elliptical dots

The provisional conclusions achieved above for the analysis of circular dots have been further confirmed and clarified by the BLS investigation of the elliptical dots, where it was possible to compare the results obtained with the exchanged SW wavevector ***k*** parallel to either the long-axis of the ellipses or their short-axis. As seen in Fig. 3, central panels, also in the case of the elliptical dots we could detect only one relatively broad peak for each spectrum and the measured frequency asymmetry $\Delta f$ was largely reduced when the lateral confinement of the nanoelement, in the direction of ***k***, was smaller than 400 nm. The latter effect is particularly evident comparing the two panels of Fig. 6. When the SW wavevector ***k*** was parallel to the minor axis of the ellipses *b*, the slope of the regression line (that according to Eq. 4 would be directly proportional to $D_{eff}$) is always appreciably smaller than in the extended film (Fig. 6(b)). Instead, when ***k*** was along the major-axis *a*, the slope increases for all the samples and that of the extended film is recovered for *a*=400 nm (Fig. 6(a)).

### B3) Stripes

The final step of our investigation was devoted to the analysis of three samples consisting of long stripes having finite widths *w*=100, 200 and 400 nm. In this case, the spatial confinement is only in one direction (rather than bi-dimensional as for the circular or elliptical dots) and we analyzed the non-reciprocity of spin waves propagating across such a finite dimension. As seen in Fig. 7(a,b,c), in this case the BLS spectra are somehow structured and we could distinguish two different modes on the Stokes side. In particular, for the narrower stripes with nominal width *w*=100 nm, we observed two modes at fixed frequencies, about 2 GHz apart from each other. The one at lower frequency has a larger cross section (full squares in Fig 7(d)) and was detected in the whole wavevector range, while the higher frequency one was detected only for positive wavenumber, with a relatively low cross section (open squares in Fig 7(d)). These two modes could be quantitatively reproduced by the micromagnetic simulations, although we had to assume an effective width of 80 nm in order to fit the calculated frequencies to the experimental points. It is clear that in this case the lateral confinement is so important that there is no chance of determining the i-DMI constants from the measured frequencies. Instead, for $w$=200 nm (Fig. 7(b) and (e)), in addition to the low-lying mode at fixed frequency, we could also measure a dispersive peak, whose frequencies are slightly lower than the dispersion curve of the extended film. Finally, for $w$=400 nm (Fig. 7(c and f)) the low-frequency mode becomes very weak, while the dominating peak follows the evolution of the Damon-Eshbach mode of the extended film, similarly to the unique peak that we could detect for the circular or the elliptical dots, because for a lateral confinement of 400 nm the eigenmodes frequencies become more dense and different flat modes contribute to the experimental peak (Fig. 7(f)).

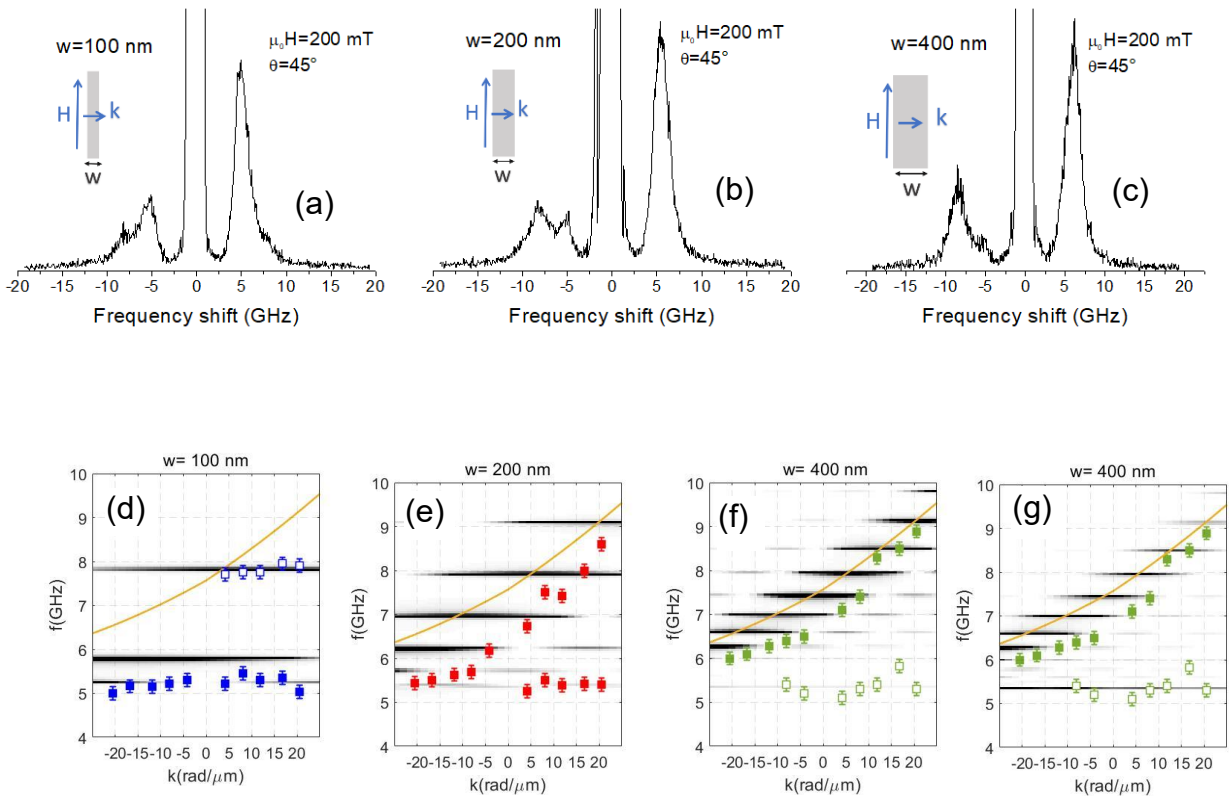

**Fig. 7** Typical BLS spectra recorded for an angle of incidence $\theta_i$=45° from the three samples consisting of long stripes with width (a) $w$=100 nm, (b) 200 nm and (c) 400 nm. The external field H is applied along the stripes, while the spin wave wavevector k is perpendicular to the stripes, as shown by the insets. Panels (d), (e) and (f) show the measured (squares) and the simulated (grey-scale curves) eigenmode frequencies for the three samples. Open squares correspond to modes with a relatively small intensity in the measured BLS spectra. It can be seen that the simulated eigenmodes are more dense in frequency as the stripe width is increased from 100 nm to 400 nm, while the extension in the k space of each mode tends to be reduced and concentrated close to the dispersion curve of the extended film (yellow line), reflecting the conservation of momentum that becomes effective as the lateral dimension of the stripes becomes comparable with the SW wavelength. For the sake of comparison, we also show

in panel (g) how the relative intensity of the simulated frequencies of the sample with *w*=400 nm changes, making the low-lying mode well visible in a large k interval, when the excitation area in the micromagnetic simulation is restricted to the edge of the wires, as explained in the main text.

One final point that deserves a short discussion has to do with the fact that in the simulation the intensity of the low-lying mode at about 5.3 GHz is very weak in the simulations, especially for the larger stripes with w=400 nm (Fig. 7f). We have verified that this is connected with the characteristics of the exciting pulse exploited in the simulations. In particular, looking at panel (g) of Fig.7 one can easily see that the low-lying mode becomes more intense in the simulations if the excitation pulse is restricted to a region of 5 nm at the edge of the stripes (rather than to a region having the width of one third of the stripe width, as in previous simulations). This preferential localization of the low-frequency mode at the edges of the stripe is a typical effect due to the presence of a sizeable DMI. In fact, it was shown in a previous theoretical study [33] that while the usual fundamental mode of a magnetic stripe uniformly magnetized along its easy axis, that appears at the bottom of the eigenmodes spectrum, has its maximum amplitude at the center of the stripe, the presence of DMI induces a strong modification of its character. In particular, it causes a decrease of its frequency and a preferential localization at the stripe edges, where the configuration of the static magnetization starts to become chiral, so that the effective field acting on the precessing spins becomes weaker and the mode amplitude increases. One may also notice in Fig. 7(g) that the strong localization of the low-lying mode in a narrow region of the real space, corresponds to an intensity distribution that extends over a relatively large interval of the *k*-space, as found in the experiment.

## V. CONCLUSIONS

Wavevector-resolved Brillouin Light Scattering experiments and micromagnetic simulations based on the GPU-accelerated MuMax3 software, have been exploited to analyze the effect of the interfacial Dzyaloshinskii–Moriya interaction on the spin wave eigenmodes of arrays of non-interacting circular and elliptical dots, as well as long stripes. These nanostructures, with lateral dimensions ranging from 100 to 400 nm, i.e. comparable with the wavelength of the thermal spin waves detected in BLS experiments, have been patterned by electron beam lithography, starting from a Pt(3.4 nm)/CoFeB(0.8 nm) bilayer. The aim of the experiments was to determine how the lateral confinement influences the spin-wave non-reciprocity induced by the presence of the sizeable i-DMI supplied by the heavy-metal substrate (Pt) on the thin ferromagnetic film (CoFeB). Our experimental results, assisted by micromagnetic simulations, provide evidence for a strong suppression of the frequency asymmetry $\Delta f$ between counter-propagating spin waves (corresponding to either Stokes or anti-Stokes peaks in BLS spectra), when the dot diameter is reduced from 400 nm to 100 nm. Such an evolution reflects the modification of the spin-wave character from propagating to stationary and indicates that the method of quantifying the i-DMI strength from the frequency difference of counter-propagating spin waves is not applicable in the case of sufficiently small magnetic elements.

## ACKNOWLEDGEMENTS

Support by the following entities is kindly acknowledged: Italian national project IT-SPIN (PRIN-2020LWPKH7) and NextGenerationEU National Innovation Ecosystem grant ECS00000041 – VITALITY, under the Italian Ministry of University and Research (MUR); EU Horizon 2020 research and innovation program, Marie Skłodowska-Curie Grant Agreement No. 860060 "Magnetism and the effect of Electric Field" (MagnEFi); DFG (CRC TRR 173 SPIN+X, #268565370, proj. A01 and B02). DFG, German Research Foundation, under Germany's Excellence Strategy – EXC-2123 QuantumFrontiers – 390837967 and within SPP 2137.

## REFERENCES

[1] Sang-Wook Cheong and Xianghan Xu, Magnetic chirality, npj Quantum Materials **7**, 40 (2022)

[2] I. Dzyaloshinsky, A thermodynamic theory of "weak" ferromagnetism of antiferromagnetics, J. Phys. Chem. Solids **4**, 241 (1958).

[3] T. Moriya, New Mechanism of Anisotropic Superexchange Interaction, Phys. Rev. Lett. **4**, 228 (1960).

[4] A. R. Fert, Magnetic and Transport Properties of Metallic Multilayers. Metallic Multilayers, in *Materials Science Forum (Vol. 59-60);* pp 439–480 (1991).

[5] A. Crépieux, C. Lacroix, Dzyaloshinsky–Moriya interactions induced by symmetry breaking at a surface, J. Magn. Magn. Mater. **182**, 341 (1998).

[6] H. Yang, A. Thiaville, S. Rohart, A. Fert, M. Chshiev, Anatomy of Dzyaloshinskii-Moriya Interaction at Co/Pt Interfaces, Phys. Rev. Lett. **115**, 267210 (2015).

[7] A. Belabbes, G. Bihlmayer, F. Bechstedt, S. Blügel, A. Manchon, Hund's RuleDriven Dzyaloshinskii-Moriya Interaction at $3d−5d$ Interfaces, Phys. Rev. Lett. **117**, 247202 (2016).

[8] L. Udvardi, L. Szunyogh, Chiral Asymmetry of the Spin-Wave Spectra in Ultrathin Magnetic Films, Phys. Rev. Lett. **102**, 207204 (2009).

[9] D. Cortés-Ortuño, P. Landeros, Influence of the Dzyaloshinskii–Moriya interaction on the spin-wave spectra of thin films, J. Phys: Cond. Matt. **25**, 156001 (2013).

[10] J.-H. Moon, S.-M Seo, K.-J. Lee, K.-W. Kim, J. Ryu, H.-W Lee, R. D. McMichael, M. D. Stiles, Spin-wave propagation in the presence of interfacial Dzyaloshinskii-Moriya interaction, Phys. Rev. B **88**, 184404 (2013).

[11] M. Kostylev, Interface boundary conditions for dynamic magnetization and spin wave dynamics in a ferromagnetic layer with the interface Dzyaloshinskii-Moriya interaction, J. Appl. Phys. **115**, 233902 (2014).

[12] A complete and updated review about the DMI determination in thin films by BLS and by other methods, together with a comprehensive bibliography, can be found in: M. Kuepferling, A. Casiraghi, G. Soares, G. Durin, F. Garcia-Sanchez, L. Chen and C. H. Back, C. H. Marrows, S. Tacchi and G. Carlotti, Measuring interfacial Dzyaloshinskii-Moriya interaction in ultra-thin magnetic films, Rev. Mod. Phys. **95**, 015003 (2023)

[13] K. Di, V.L. Zhang, H.S. Lim, S. C. Ng, M.H. Kuok, J. Yu, J. Yoon, X. Qiu, H. Yang, Direct observation of the Dzyaloshinskii-Moriya interaction in a Pt/Co/Ni film, Phys. Rev. Lett. **114**, 047201 (2015).

[14] H.T. Nembach, J.M. Shaw, M. Weiler, E. Ju'e, T.J. Silva, Linear relation between Heisenberg exchange and interfacial Dzyaloshinskii–Moriya interaction in metal films, Nat. Phys. **11**, 825 (2015).

[15] A.K. Chaurasiya, C. Banerjee, S. Pan, S. Sahoo, S. Choudhury, J. Sinha, A. Barman, Direct Observation of Interfacial Dzyaloshinskii-Moriya Interaction from Asymmetric Spin-wave Propagation in W/CoFeB/SiO$_2$ Heterostructures Down to Sub-nanometer CoFeB Thickness, Sci. Rep. **6**, 32592 (2016).

[16] S. Tacchi, R.E. Troncoso, M. Ahlberg, G. Gubbiotti, M. Madami, J. Åkerman, P. Landeros, Interfacial Dzyaloshinskii-Moriya Interaction in Pt/CoFeB Films: Effect of the Heavy-Metal Thickness, Phys. Rev. Lett. **118**, 147201, (2017).

[17] M. Belmeguenai, M.S. Gabor, Y. Roussigné, T. Petrisor, R.B, Mos, A. Stashkevich, S.M. Chérif, C. Tiusan, Interfacial Dzyaloshinskii-Moriya interaction sign in Ir/Co2FeAl systems investigated by Brillouin light scattering, Phys. Rev. **B 97**, 054425 (2018).

[18] H. Bouloussa, J. Yu, Y. Roussigné, M. Belmeguenai, A. Stashkevitch, H. Yang, S.M. Cherif, B. Mohamed, A. Stachkevich, Brillouin light scattering investigation of interfacial Dzyaloshinskii–Moriya interaction in ultrathin Co/Pt nanostripe arrays, J. Phys. D: Appl. Phys. **51**, 225005 (2018).

[19] Y. Chen, Q. Zhang, J. Jia, Y. Zheng, Y. Wang, X. Fan, J. Cao, Tuning Slonczewski-like torque and Dzyaloshinskii–Moriya interaction by inserting a Pt spacer layer in Ta/CoFeB/MgO structures, Appl. Phys. Lett. **112**, 232402 (2018).

[20] G.W Kim, A.S. Samardak,; Y.J. Kim,; I.H. Cha, A.V. Ognev, A.V. Sadovnikov, S.A.Nikitov, Y.K. Kim, Role of the Heavy Metal's Crystal Phase in Oscillations of Perpendicular Magnetic Anisotropy and the Interfacial Dzyaloshinskii-Moriya Interaction in W/Co−Fe−B/MgO Films, Phys. Rev. Appl. **9**, 064005 (2018).

[21] J. Lan, W. Yu, R. Wu, J. Xiao, Spin-Wave Diode, Phys. Rev. X **5**, 041049 (2015).

[22] J.-V Kim, R.L. Stamps, R.E. Camley, Spin Wave Power Flow and Caustics in Ultrathin Ferromagnets with the Dzyaloshinskii-Moriya Interaction, Phys. Rev. Lett. **117**, 197204 (2016).

[23] T. Brächer, O. Boulle, G. Gaudin, P. Pirro, Creation of unidirectional spin-wave emitters by utilizing interfacial Dzyaloshinskii-Moriya interaction, Phys. Rev. B **95**, 064429 (2017).

[24] K. Szulc, P. Graczyk, M. Mruczkiewicz, G. Gubbiotti, M. Krawczyk, Spin-WaveDiode and Circulator Based on Unidirectional Coupling, Phys. Rev. Applied, **14**, 034063 (2020).

[25] J. Chen, H. Yu, G. Gubbiotti, Unidirectional spin-wave propagation and devices, J. Phys. D: Appl. Phys **55**, 123001 (2021).

[26] S.-J Lee, J.-H Moon, H.-W Lee, K.-J. Lee, Spin-wave propagation in the presence of inhomogeneous Dzyaloshinskii-Moriya interactions, Phys. Rev. B **96**, 184433 (2017).

[27] R.A. Gallardo, D. Cortés-Ortuño, T. Schneider, A. Roldán-Molina, F. Ma, R.E. Troncoso, K. Lenz, H. Fangohr, J. Lindner, P. Landeros, Flat Bands, Indirect Gaps, and Unconventional Spin-Wave Behavior Induced by a Periodic Dzyaloshinskii-Moriya Interaction, Phys. Rev. Lett. **122**, 067204 (2019).

[28] R.A. Gallardo, D. Cortés-Ortuño, R.E. Troncoso, P. Landeros, In Three-Dimensional Magnonics: Layered, Micro- and Nanostructures; Gubbiotti, G., Ed.; Jenny Stanford Publishing: Berlin, Heidelberg, 2019; pp 121–160.

[29] R. Silvani, M. Kuepferling, S. Tacchi and G. Carlotti, Impact of the interfacial Dzyaloshinskii-Moriya interaction on the band structure of one-dimensional artificial magnonic crystals: a micromagnetic study, J. Magn. Magn. Mater. **539**, 168342 (2021)

[30] F. Garcia-Sanchez, P. Borys, A. Vansteenkiste, J.V. Kim, R.L. Stamps, Nonreciprocal spin-wave channeling along textures driven by the Dzyaloshinskii-Moriya interaction, Phys Rev. B, **89**, 224408 (2014).

[31] M. Mruczkiewicz, M. Krawczyk, Influence of the Dzyaloshinskii-Moriya interaction on the FMR spectrum of magnonic crystals and confined structures, Phys. Rev. B, **94**, 024434 (2016).

[32] B.W. Zingsem, M. Farle, R.L. Stamps, R.E. Camley, Unusual nature of confined modes in a chiral system: Directional transport in standing waves, Phys. Rev. B, **99**, 214429 (2019).

[33] R. Silvani, M. Alunni, S. Tacchi and G. Carlotti, Effect of the Interfacial Dzyaloshinskii–Moriya Interaction on the Spin Waves Eigenmodes of Isolated Stripes and Dots Magnetized In-Plane: A Micromagnetic Study, Appl. Sci **11**, 2929 (2021)

[34] A. Vansteenkiste, J. Leliaert, M. Dvornik, M. Helsen, F. Garcia-Sanchez, B. Van Waeyenberge, The design and verification of MuMax3, AIP Adv. **4**, 107133 (2014)

[35] R. Zivieri, F. Montoncello, L. Giovannini, F. Nizzoli, S. Tacchi, M. Madami, G. Gubbiotti, G. Carlotti, and A. O. Adeyeye, Collective spin modes in chains of dipolarly interacting rectangular magnetic dots, Phys. Rev. **B 83**, 054431 (2011)

[36] Kai Di, Vanessa Li Zhang, Hock Siah Lim, Ser Choon Ng, Meng Hau Kuok, Jiawei Yu, Jungbum Yoon, Xuepeng Qiu, and Hyunsoo Yang. Direct observation of the Dzyaloshinskii-Moriya interaction in a Pt/Co/Ni film, Phys. Rev. Lett. **114**, 047201 (2015)

[37] Avinash Kumar Chaurasiya, Chandrima Banerjee, Santanu Pan, Sourav Sahoo, Samiran Choudhury, Jaivardhan Sinha & Anjan Barman, Direct Observation of Interfacial Dzyaloshinskii-Moriya Interaction from Asymmetric Spin-wave Propagation in W/CoFeB/SiO2 Heterostructures Down to Subnanometer CoFeB Thickness, Sci Rep **6**, 32592 (2016).